\documentclass[a4paper,12pt]{article}
\usepackage[utf8]{inputenc}
\usepackage[english]{babel}
\usepackage{authblk}
\usepackage{graphicx} 
\usepackage{amsmath}
\usepackage{mathptmx}
\usepackage[singlespacing]{setspace}
\usepackage[headheight=1in,margin=1in]{geometry}
\usepackage{fancyhdr}

\usepackage{amsfonts}

\usepackage{hyperref}

\usepackage[
    maxcitenames=1
]{biblatex} 
\makeatletter
\def\@maketitle{%
  \newpage

  \begin{center}%
  \let \footnote \thanks
    {\LARGE \@title \par}%
  \end{center}%
  \par
  \vskip 0.1em}
\makeatother

\title{Conditioning LLMs on Social Value Orientation improves behavioural alignment in a sequential social dilemma}

\date{}

\begin{document}

\maketitle

\begin{center}
    Marco Saponara$^{1,2,3}$, 
    Axel Abels$^{1,2,3,4}$,
    Ann Nowé$^{2,3}$,
    Tom Lenaerts$^{1,2,3,5}$\\
	\small$^{1}$Machine Learning Group, Universit\'e Libre de Bruxelles, Brussels, Belgium\\
	\small$^{2}$Artificial Intelligence Lab, Vrije Universiteit Brussel, Brussels, Belgium\\
    \small$^{3}$FARI Institute, Universit\'e Libre de Bruxelles and Vrije Universiteit Brussel, Brussels, Belgium\\
    \small$^{4}$ELLIS Alicante, Alicante, Spain\\
    \small$^{5}$Center for Human-Compatible AI, UC Berkeley, Berkeley, CA, USA
\end{center}

\begin{abstract}
    Large language Models (LLMs) are increasingly used to simulate human decision-making, yet their outputs often under-represent human behavioural heterogeneity. We investigate whether conditioning LLMs on Social Value Orientation (SVO, a measure of how individuals value their own outcomes relative to others’) can better reproduce human behaviour in a sequential social dilemma. Using experimental data from two variants of the Centipede Game (CG) as reference, we compare the default behaviour of eight LLMs with behaviour generated after conditioning them on SVO profiles drawn from the human sample. We find that the default strategies vary substantially across models, but are generally distant from the reference distributions. However, conditioning them on SVO profiles systematically steers their strategic behaviour, improving alignment with the human reference by up to 70\% with respect to the default. Across models, higher induced SVO values decrease the probability of stopping the game, reproducing the relationship between prosociality and cooperation observed in human behaviour. Together with the sensitivity of LLMs' elicited SVO to prompt and order effects, these results suggest  that the usefulness of SVO for behavioural simulation does not depend on LLMs possessing stable social preferences, but rather on their ability to map social preferences onto corresponding strategic choices.
\end{abstract}


\section{Introduction} 

Since the public release of ChatGPT in 2022, the use of Large Language Models (LLMs) has dramatically surged.
Although these models operate by predicting the next token in a sequence given a specific context (or \emph{prompt}), their textual output has become virtually indistinguishable from humans~\cite{doi:10.1073/pnas.2524472123}.
Notably, recent works have argued that LLMs appear to simulate complex features of human cognition, such as theory of mind~\cite{strachan2024testing, kosinski2024evaluating}, emotional intelligence~\cite{schlegel2025large}, and even deceptive mechanisms~\cite{hagendorff2024deception}.

For these reasons, even though LLMs are neither sentient nor intentional, an increasing corpus of research is investigating their capabilities as strategic decision-makers in both individual and interactive tasks~\cite{hutson2023guinea, bail2024can, akata2025playing}.
Remarkably, LLMs appear to be capable of predicting the outcome of social science experiments and show signs of social preferences, reciprocity, and inference of the beliefs and intentions of others in economic contexts~\cite{ashokkumar2026large, horton2023large}.
On the other hand, current concerns include the LLMs’ tendency to understate the diversity of human responses and to exhibit systematic biases toward groups that are inaccurately or inadequately represented in their training data \cite{abels2026improving, wang2025large}.
In addition, results may vary depending on factors such as task-specific setting, framing, prompt design, and model configuration \cite{schmidt2024gpt, sclar2024quantifying}.
Thus, it remains unclear to what extent these systems can constitute a robust proxy for human samples in behavioural and social sciences~\cite{gao2025take}.

A potential avenue to address these issues consists in explicitly conditioning the model on certain variables such as psychological traits and demographics~\cite{argyle2023out}.
Indeed, if prior information on the population-level distribution of these variables is available, it may be possible to construct a population of LLM-based agents representing distinct \emph{personas}, thus approximating the heterogeneity observed in human samples.
For example, recent research~\cite{sakai2026effects} has explored how certain personality traits, measured through the Big Five framework, can influence cooperation in a social dilemma.
The study assessed LLMs' personality scores, examined strategic behaviour with and without explicit personality information, and tested the effects of independently setting each personality dimension to extreme values.
Here, we go beyond this methodology by asking not only (1) how the decision-making of an LLM can be steered to obtain a diverse range of strategic behaviours, but also (2) in which cases they can align with a known human sample, through an analysis grounded on experimental data.

\begin{figure}[!t]
\centering
\resizebox{\textwidth}{!}{%
\includegraphics{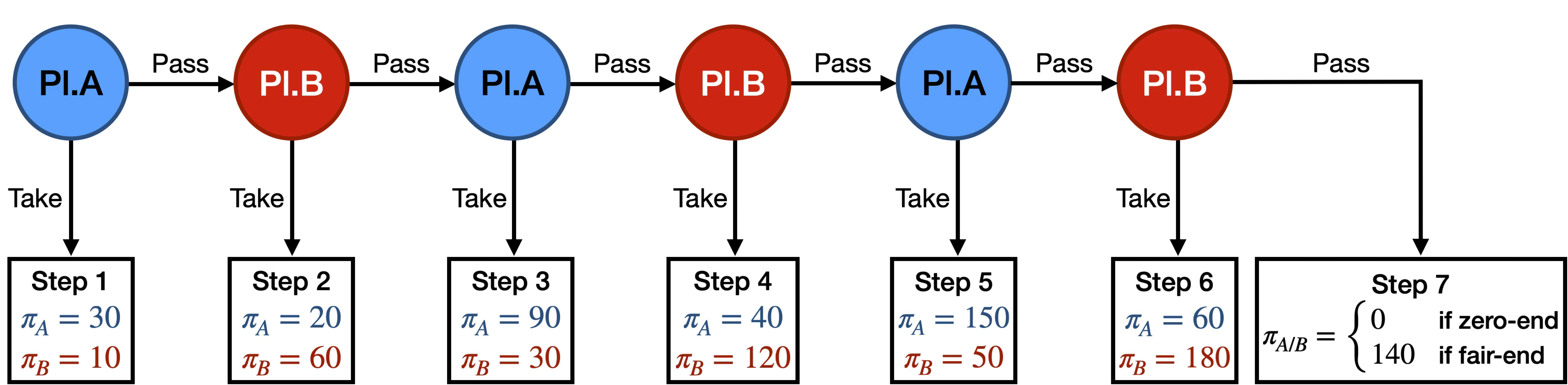}
}
\caption{Extensive form of the two Centipede Games used in our experiment.
Player A (B) plays at odd (even) steps. 
The shared resource starts at 40 and increases linearly by 40 units for six turns. 
Playing \emph{Take} at a given step means ending the game and receiving 75\% of the resource, while the co-player receives the remaining 25\%.
Playing \emph{Pass} means letting the other player decide what to do in the next step, unless the last decision node is reached, where Player B must decide between two different splits. 
We propose two types of final split:
in the zero-end game both players receive 0, whereas in the fair-end game both players receive 50\% of the resource.
}
\label{fig:1}
\end{figure}

The present study builds on a previous behavioural experiment~\cite{saponara2026prospect} about a sequential social dilemma known as the Centipede Game (CG)~\cite{rosenthal_games_1981}.
In this task, two players take turns to decide how to split a shared resource (see \autoref{fig:1} and \autoref{sec:methods}).
At each decision node, the focal player can either stop the interaction and receive a larger share of the resource (action \emph{Take}) or proceed to the following turn and pass this decision to the co-player (action \emph{Pass}).
The dilemma arises particularly when the shared resource increases with each step, forcing each player to choose between one’s immediate personal gain and the greater long-term benefit available to both players.
Since the CG describes turn-based mixed-motive interactions, it is particularly relevant in the context of LLM-based research.
Indeed, these systems are frequently used as conversational chatbots, and users engage in repeated exchanges with them to achieve a personal goal.
However, prompts often entail a cost for the user, while the LLM consumes varying resources to generate responses that may help the user reach their goal.
The sequential nature of the conversation, along with its costs and its uncertain gains, make the CG a suitable approximation for these interactions.

The behavioural experiment on which the present analysis is based compared the outcomes of two under-explored variations of a linearly-incremental CG, namely the \emph{zero-end} CG (where the final step is a deadline leading to zero payoffs for both players, $n=164$) and the \emph{fair-end} CG (where the last split is a fair division of the shared resource, $n=162$)~\cite{saponara2026prospect}.
The findings suggest that the prospect of a fair outcome, even though uncertain, prompts subjects to go beyond self-interested behaviour. 
Furthermore, cooperative behaviour in the CG is correlated with individuals’ prosociality, measured through the Social Value Orientation (SVO)~\cite{murphy2011measuring}, a resource allocation task that measures how much weight a person gives to the welfare of others in relation to their own, thus corroborating the idea that social preferences constitute an important driver of human behaviour in this type of social dilemmas.

Although there is broad consensus that SVO is associated with cooperation in social dilemmas because it captures how individuals subjectively value outcomes, it has not yet been applied in the context of the conditioning and alignment of LLMs in strategic interactions~\cite{pletzer2018social, balliet2009social}.
The present study aims to bridge this gap by investigating whether inducing a particular SVO profile in an LLM can steer its behavioural choices in the CG accordingly. 
Specifically, our objective is the comparison between the default behaviour of different LLMs playing the CG without any conditioning and the behaviour obtained after prompting the LLM with several SVO profiles retrieved from the experimental data. 
We expect that, in the latter case, the outcome distribution at the population level, obtained by aggregating all behavioural profiles according to the prevalence of their respective SVO values in human data, is more closely aligned with the diverse human reference than the default distribution. 
In particular, we hypothesise that this increased diversity stems from the fact that a prosocial conditioning leads to higher-payoff outcomes in the CG, while an individualistic conditioning shifts the LLM’s behaviour towards the rational but sub-optimal outcome in the CG.

We tested eight LLMs (four closed models of the GPT family, two versions of Qwen-3.5, Llama-3.3 and Gemma-4).
All models, used through the Python library \emph{litellm}\footnote{\href{https://www.litellm.ai}{https://www.litellm.ai}}, were employed without reasoning and all results are based on the top-5 log-probabilities of the first generated token (see \autoref{sec:methods} for more details).
Our prompts fall into two main categories, labelled as \emph{blinded} and \emph{unblinded}~\cite{gui2023challenge}.
In the blinded system prompt the LLM is told to be \textit{a participant in a behavioural experiment}, and to choose the action it prefers the most.
In the unblinded system prompt, instead, we make the objective of our analysis explicit by imposing the LLM to be \textit{an expert in predicting human decision-making in behavioural experiments}, and to predict the most preferred choice by a participant in our experiment (all prompts are included in \autoref{appendix:prompts}).
This choice is motivated by the fact that, while \emph{blinding}\footnote{The standard practice of withholding certain information that could influence participants until the experiment is completed.} prevents experimental design from influencing human participants’ decisions, it may create confounds in LLM simulations by forcing the model to infer missing context differently across treatment conditions~\cite{gui2023challenge}.

Our contributions are the following.
We analyse the LLMs' default behaviour at each stage of both the zero- and fair-end CGs.
We find a diverse range of default behaviours:
some models follow rational self-interested strategies, while others deviate from the latter by displaying various degrees of cooperation.
Notably, these behaviours remain relatively stable across game and prompt types.
However, a common feature of these strategies is the higher degree of determinism compared to the human reference. 
The SVO conditioning procedure allows us to partially overcome this limitation.
Indeed, through SVO conditioning, the LLM-generated output systematically gets closer to the human experimental reference relative to the default behaviour in the CGs, with Qwen3.5-27B and GPT-5.4 Mini achieving the best fit in the zero- and fair-end CGs, respectively.
The overall improvement is particularly relevant in the fair-end case.
However, many models preserve their characteristic behaviour, often remaining further from the human distribution than the baseline threshold set by a random agent that selects each action with equal probability.
The improvement obtained through SVO conditioning is due to the ability of LLMs to map the induced SVO profile to strategic choices in the CG.
Indeed, we find that higher SVO values (i.e., more prosocial profiles) lead to a lower probability of stopping at each step of the game.
Nevertheless, when SVO in LLMs is elicited through the original task, results appear to sensitive to the prompt type and subject to order bias, suggesting that their SVOs are closer to task-conditioned profiles than to stable human-like social preferences.

\section{Results and discussion}

\subsection*{Default behaviour in the CG varies by model but remains stable across game and prompt types.}

We start our analysis by examining the default behaviour of LLMs in both the zero- and fair-end CGs.
We elicit behaviour by prompting the LLM to choose an action at each decision node, either \emph{Take} (labelled \emph{X}) or \emph{Pass} (labelled \emph{Y}), following a similar approach as the strategy method developed by Selten~\cite{selten2003play}.
This approach has the advantage of avoiding costly simulations as well as having a complete picture of the strategic profile of the model at every stage of the game.
The behaviour in the CG is then defined by the probability of choosing to stop the game at each decision node $n\in\left\{1,\ldots, 6\right\}$ given that such node is reached, $P(Take|n)\in[0,1]$. 
This probability is directly obtained from the model's log-probability for the token corresponding to the action \emph{Take}, conditioned on the prompt presented at node $n$ (see \autoref{appendix:prompts}).

Each Panel in \autoref{fig:cg-baseline} shows, for a given LLM, the estimated \emph{Take} probability as a function of the decision node, separately for game (zero/fair-end) and prompt types (blinded/unblinded, see \autoref{appendix:prompts}).
While LLMs display heterogeneous behaviour, even within the same family, variation in behaviour appears to be driven more strongly by the choice of LLM than by game and prompt type.
These results suggest that models such as GPT-5.4-Nano, Qwen3.5-9B, Gemma-4-31B follow strategies characterised by a high degree of game-theoretical rationality: they choose \emph{Take} almost deterministically in the majority of decision nodes.
In contrast, for GPT-5.4-Mini and Llama-3.3-70B  the probability of choosing \emph{Take} drops in the fourth decision node of both games, as well as in the sixth node of the fair-end CG.
These models thus appear to be sensitive to their role in the game, as the \emph{Take} probability is on average lower for Player B on even nodes.
Notably, GPT-5.1 is the only model that exhibits a more cooperative behaviour, with a deterministic \emph{always-Pass} strategy in the fair-end game and a \emph{Take} probability which only increases sharply in the second half of the zero-end game.
Interestingly, this cooperation is not maintained by GPT-5.4 (released less than six months after GPT-5.1), as it chooses \emph{Take} deterministically starting from the third decision node.
One of the potential reasons for this sudden change between incremental improvements of OpenAI's frontier model could be a correction for \emph{sycophancy}\footnote{In AI research, sycophancy describes the tendency of LLMs to shape their responses around what they infer users would prefer to read or hear rather than the accurate or warranted response.}.

The results appear to be robust to minor prompt variations (see \autoref{fig:cg-baseline-invrole} and \autoref{fig:cg-baseline-invact} in \autoref{appendix:sensitivity}).
However, it is important to mention that the models achieve varying degrees of success rate on the comprehension test taken from the original behavioural experiment (see \autoref{fig:comprehension} in \autoref{appendix:comprehension}):
While the majority reach a success probability higher than 97\%, GPT-5.4-Mini, GPT-5.4-Nano, and Qwen-3.5-9B have an overall success probability lower than 23\%, with the first two failing almost deterministically at the fourth question of the test.
Nevertheless, we do not exclude these models from the analysis, as they might be capable of enacting meaningful behavioural changes based on external variables.
Finally, the relatively high degree of determinism in the default behaviour, as well as the fact that the choices of most LLMs are not affected by the game type suggest that, in this context, LLMs remain unsuitable for hypothesis generation and pilot studies. 
To overcome this issue, the following section explores whether SVO conditioning can bring LLM generated responses closer to human behaviour, and thus more useful for behavioural sciences.

\begin{figure}[!ht]
\centering
\resizebox{\textwidth}{!}{%
\includegraphics{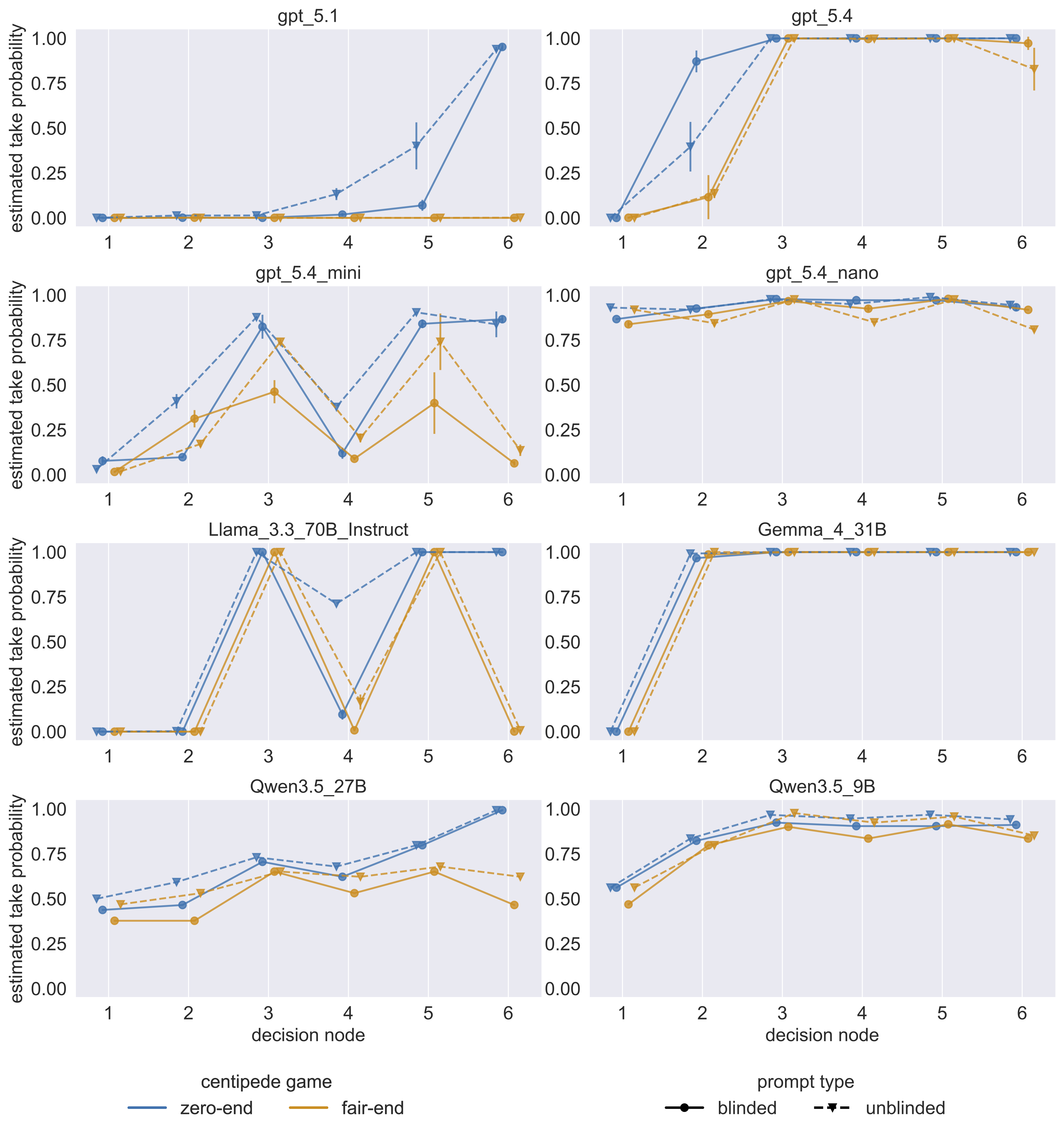}
}
\caption{Default behaviour of LLMs in the CG.
The Panels show the estimated probability of choosing \emph{Take} at each decision node of the game.
Each Panel is associated with a different LLM.
The colour of the lines specifies the type of CG (zero-end or fair-end), while the line and marker styles determines the prompt type (blinded or unblinded).
Each point is the average of 9 generations of the same probability value with different seeds.
The vertical bars represent the standard deviation of the estimates.
}
\label{fig:cg-baseline}
\end{figure}


\subsection*{SVO conditioning drives LLMs towards more representative behaviour in the CG.}

Having established the default behaviour of eight LLMs in the CGs, we now investigate the effects of SVO conditioning on the behavioural alignment of these models.
As stated in the Introduction, we repeatedly condition LLMs on several SVO profiles retrieved from a human dataset and ask them to make decisions in the CGs in accordance with these profiles (see \autoref{sec:methods} for more details).
In the context of the CG, prosocial individuals tend to be less inclined to choose \emph{Take} at a given node, whereas individualists may pursue a more self-interested strategy, thereby increasing the chances of stopping the game early.
SVO conditioning thus aims to induce representative behaviour in the CG by exploiting the potential correlation between an individual's social preferences and their strategic behaviour.
If such relationships are sufficiently represented in an LLM's training data, the model may be able to reproduce them, thereby generating responses that more closely resemble human behaviour.

\begin{figure}[!ht]
\centering
\resizebox{\textwidth}{!}{%
\includegraphics{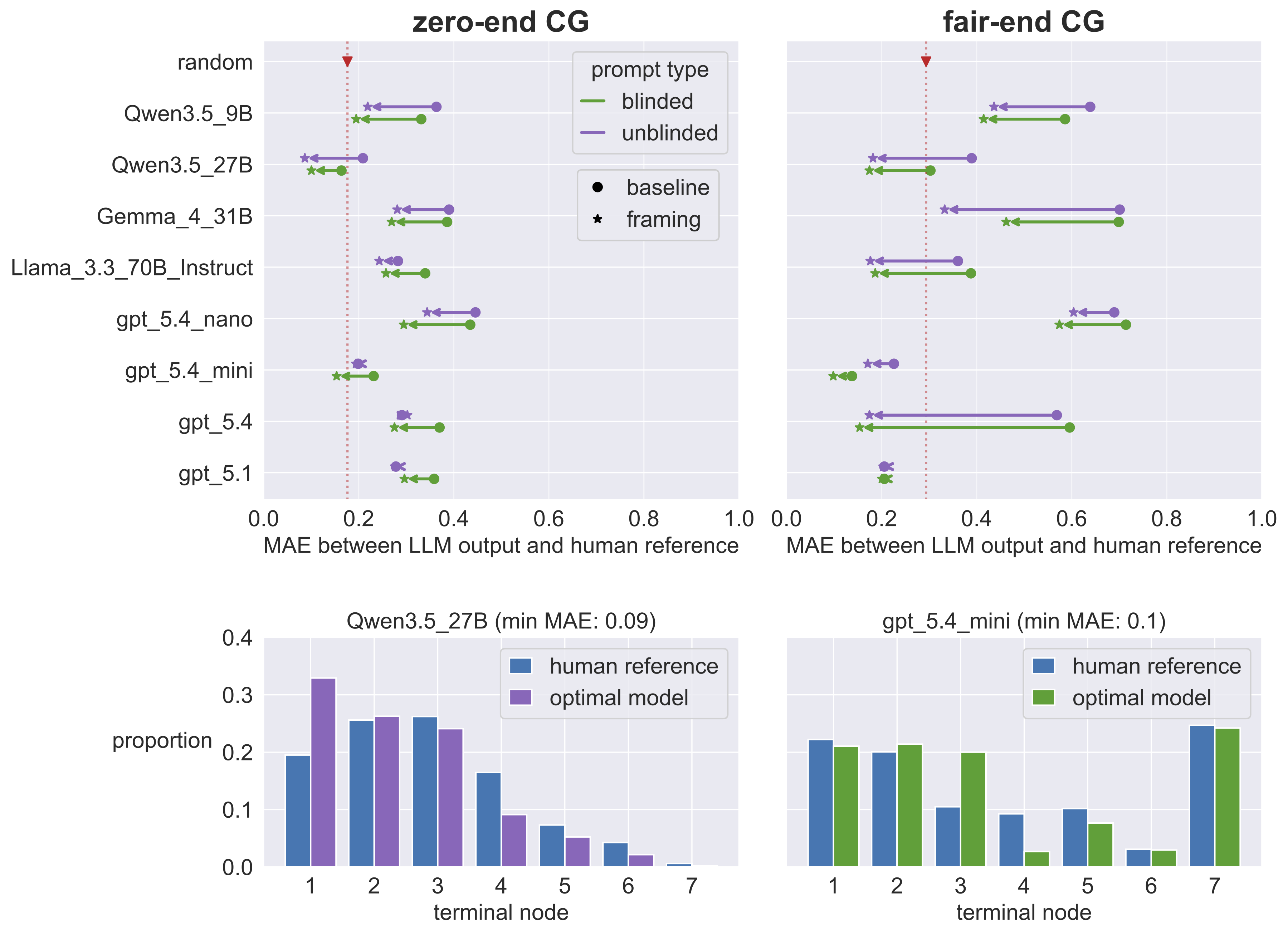}
}
\caption{Improvement in fit to the human experimental reference achieved by SVO conditioning relative to the LLMs' baseline behaviour in the CG.
The panels on the left (right) regard the zero-end (fair-end) CG.
The top panels show, for each LLM on the vertical axis, the improvement (i.e., the decrease in distance from the human reference of the behavioural experiment) as a straight arrow from the baseline to the value achieved through SVO conditioning, separately for each prompt type. 
The red dotted vertical line in the top panels represents the threshold distance set by a random agent who chooses \emph{Take} with 50\% probability at each node.
The bottom panels show the comparison between the experimental reference and the optimal LLM-generated outcome distribution.}
\label{fig:cg-conditioning}
\end{figure}

\autoref{fig:cg-conditioning} shows that SVO conditioning indeed results in a systematic reduction in mean absolute error (MAE)\footnote{Given two $n$-dimensional arrays $x$ and $y$, the mean absolute error between $x$ and $y$ is computed as $MAE(x,y) = \frac{1}{n}\sum_{i=1}^{n}|x_{i} - y_{i}|$.} with respect to the human experimental reference.
Here, the MAE is computed between the take probabilities generated by the LLMs and those estimated from the human reference data.
The improvement is particularly large in the fair-end CG (right panels), with an average relative reduction in MAE of 36\% (standard deviation $\sigma=0.21$), than in the zero-end CG (left panels), where the average relative reduction is 25\% ($\sigma=0.16$).
One possible explanation for this difference is that the LLMs' default behaviours in the fair-end game are more dissimilar to the human reference ($\mu_{baseline, fair}=0.46$, $\sigma_{baseline, fair}=0.20$) than in the zero-end game ($\mu_{baseline, zero}=0.32$, $\sigma_{baseline, zero}=0.08$), leaving greater room for improvement under SVO conditioning.
Through this procedure, the gap between game types decreases substantially ($\mu_{cond, fair}=0.28$, $\sigma_{cond, fair}=0.16$; $\mu_{cond, zero}=0.24$, $\sigma_{cond, zero}=0.07$).
The results are typically consistent across prompt variants.
Finally, for both games, the lowest MAE ($\approx 0.1)$ is obtained by SVO conditioning: Qwen3.5-27B with unblinded prompts in the zero-end CG, and GPT-5.4-Mini with blinded prompts in the fair-end CG.
The optimal terminal node distributions are compared to the human reference in the bottom panels of \autoref{fig:cg-conditioning}.

Despite the improvements achieved through SVO conditioning, in the zero-end CG, a random agent that chooses \emph{Take} with 50\% chance at each decision node more closely approximates the human distribution than all but two models, Qwen3.5-27B and GPT-5.4-Mini (see the red dotted vertical line in the top panels).
In the fair-end case, on the other hand, most models outperform the random player, with the exception of Qwen3.5-9B, Gemma-4-31B, and GPT-5.4-Nano.
This finding suggests that the utility of conditioning LLMs on social preferences is highlighted by the presence of the fair outcome of the fair-end CG.

\subsection*{The higher the conditioned SVO, the lower the chances of stopping the CG.}

Given that the improvements from SVO conditioning often exceed what would be expected from simply introducing random variation, it is likely that the resulting distributions follow a specific pattern rather than simply being random noise.
Here, we aim to quantify this pattern by assessing how strategic choices at each stage of the game depend on individual SVO profiles.
Thus, for each game and each decision node $n\in\left\{1,\ldots, 6\right\}$, we fit a Generalized Linear Model (GLM) with a binomial family and logit link to the behavioural data obtained through SVO conditioning.
Specifically, the GLM is a non-linear mapping $f_{n}$ from the scalar SVO value $x$ to the estimated probability of choosing \emph{Take} at node $n$: $y=f_{n}(x;\beta_{0}, \beta_{1}) = 1/(1+e^{-(\beta_{0} + \beta_{1}x)})$.
The two parameters $\beta_{0}$ and $\beta_{1}$ are the maximum likelihood estimates of the model, with $\beta_{0}$ determining the baseline \emph{Take} probability, and $\beta_1$ capturing how this tendency varies with SVO.

\begin{figure}[!ht]
\centering
\resizebox{\textwidth}{!}{%
\includegraphics{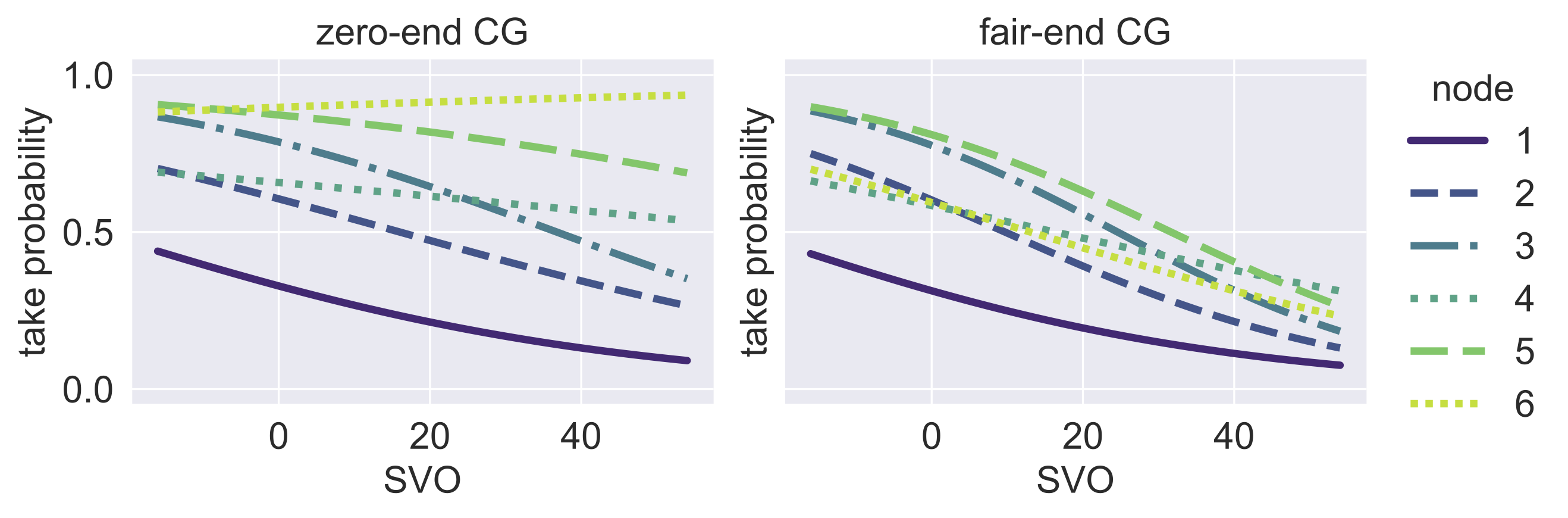}
}
\caption{Mapping between the SVO value induced through SVO conditioning and the estimated \emph{Take} probability at each decision node of the CG.
The panels show, for the zero-end (left) and fair-end (right) CGs, the estimated probability of choosing \emph{Take} as a function of different SVO values, for all decision nodes $1\leq n \leq 6$.
The lines in each panel are obtained by fitting different GLMs with binomial family and logit link on the behavioural data of all LLMs under SVO conditioning, with blinded and unblinded prompts. (See \autoref{appendix:glm-summary} for the summaries of the models.)
}
\label{fig:svo-effect}
\end{figure}

\autoref{fig:svo-effect} shows the GLM curves for each decision node, each fitted to data aggregated across all prompt types and LLMs (in \autoref{appendix:fig4extended} we show the results for each LLM separately, and in \autoref{appendix:glm-summary} we report the models' summary).
The panels show the probability of choosing \emph{Take} at a given decision node $n$ as a function of the SVO value induced by the SVO conditioning, separately for the zero- and fair-end CGs.
These results confirm that higher SVOs (i.e., the more prosocial the LLM is told to act) generally induce lower \emph{Take} probabilities. 
The only exception is the last decision node of the zero-end CG, where the trend is slightly increasing and the \emph{Take} probability stays consistently above 80\%.
However, this increase shows that prosocial agents are even more likely to avoid the zero-payoff outcome, as it would negatively affect both players, hence remaining coherent with the other nodes.
Notably, the curves associated with later decision nodes lie above the curves of earlier nodes. 
The only exceptions involve the 4th node in the zero-end CG and the 4th and 6th nodes in the fair-end CG.
This means that, excluding the aforementioned exceptions, the \emph{Take} probability increases as the game progresses regardless of the model's SVO, showing further consistency along the decision node dimension.

\subsection*{LLMs' elicited SVO is sensitive to prompt type and shows signs of order bias.}

The results of the previous section suggest that, when LLMs are explicitly conditioned on SVO, they appear to adapt their output in the CG accordingly, with prosocial profiles leading to more cooperative choices in the game.
Motivated by this effect, we elicit the LLMs' inherent social preferences to examine if they can explain the default output observed in the CGs.
We then estimate the default SVO distributions of the LLMs (that is, their SVO distributions when no SVO conditioning is applied) using the original Slider Measure developed by \textcite{murphy2011measuring}.
Specifically, each distribution is obtained by aggregating independent responses across the six primary items of the Slider Measure (see \autoref{sec:methods} for more details).
On average, however, we do not detect any effect between the elicited default SVO and the default output in the CGs (see \autoref{fig:svo-cg} in \autoref{appendix:svo-cg}). 
In this section, we further explore the limitations of using this task to draw meaningful conclusions about LLMs' inherent social preferences.
In particular, we investigate whether LLMs are robust to changes in how the SVO task is presented.
For each LLM, the SVO distribution is computed under four conditions, combining two prompt types (blinded and unblinded) with two orderings of the nine response options within each item of the task (standard and inverted). 
The comparison between blinded and unblinded prompts aims to capture sensitivity to prompt type, while reversing the original option order allows us to test for potential order bias.

\begin{figure}[!ht]
\centering
\resizebox{\textwidth}{!}{%
\includegraphics{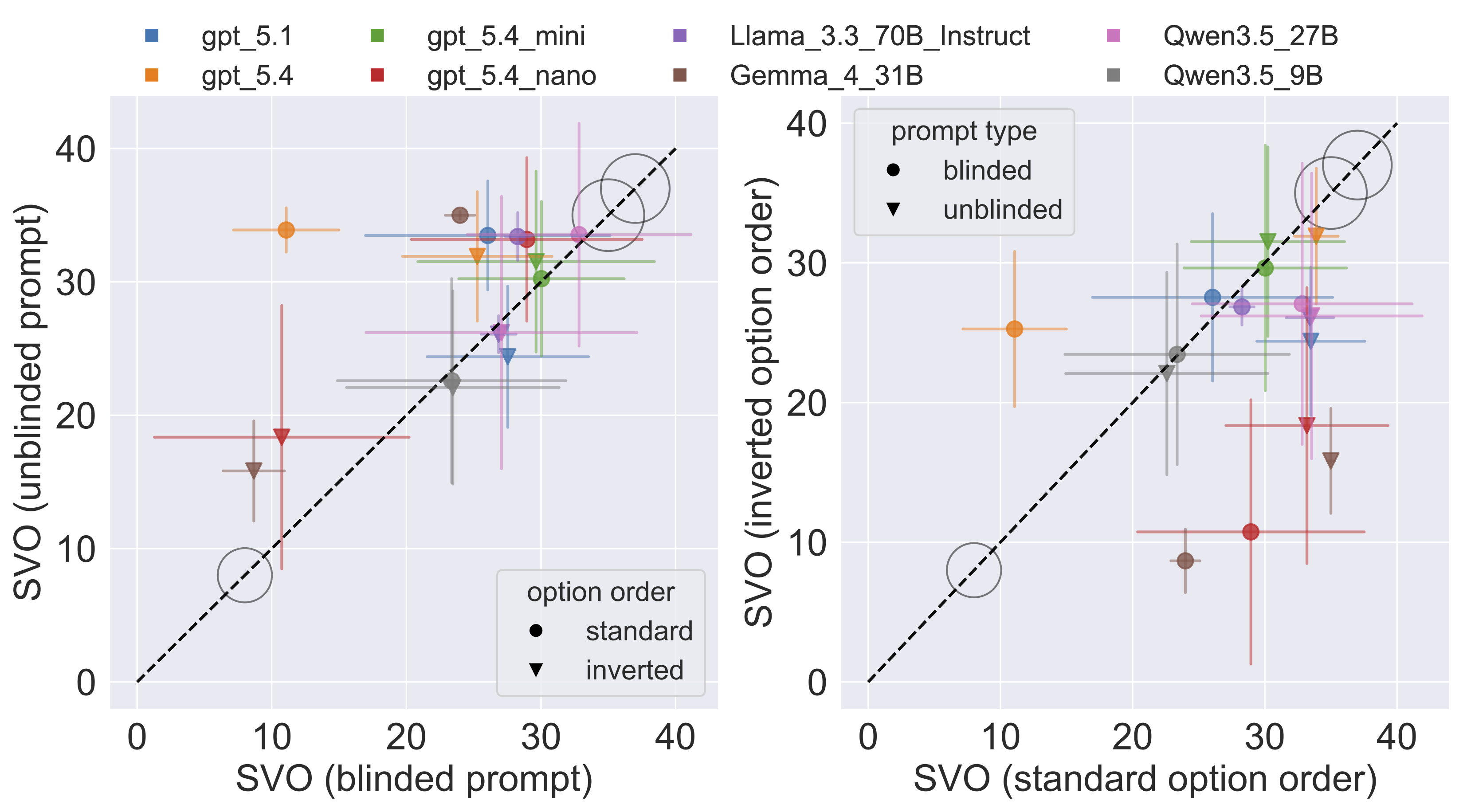}
}
\caption{Shifts in SVO-distribution means induced by prompt type and option order in the SVO task.
In the left panel, each point represents a pair of distribution means obtained using the blinded prompt (\(x\)-axis) and the unblinded prompt (\(y\)-axis), \((\mu_{\mathrm{SVO}}^{\mathrm{blinded}}, \mu_{\mathrm{SVO}}^{\mathrm{unblinded}})\). 
In the right panel, each point represents a pair of distribution means obtained using the standard option order (\(x\)-axis) and the inverted option order (\(y\)-axis), \((\mu_{\mathrm{SVO}}^{\mathrm{standard}}, \mu_{\mathrm{SVO}}^{\mathrm{inverted}})\). 
The lines indicate the standard deviations of the corresponding distributions. 
Marker colour identifies the LLM, while marker shape indicates the option order in the left panel and the prompt type in the right panel.
The three empty circles on the diagonal represent the SVO values in the human distribution having a frequency mass greater than 5\% ($\left\{8, 35, 37\right\}$, with respective frequencies $f_{8}=.09$, $f_{35}=.15$, $f_{37}=.14$).
}
\label{fig:svo}
\end{figure}

\autoref{fig:svo} reports the shifts in mean SVO induced by these two manipulations.
The left panel compares the blinded condition, in which the LLM is asked to make its preferred choice as a participant, with the unblinded condition, in which it is asked to predict the choice a participant would prefer. 
On average, SVO tends to be higher under the unblinded prompt, suggesting that the elicited social preferences depend on whether LLMs are asked to choose for themselves or predict the preferences of others. 
This tendency is not uniform across models. 
Most models exhibit relatively small shifts, whereas GPT-5.4 is particularly sensitive to unblinding, shifting from a self-interested to a prosocial SVO when options are presented in the standard order.
The right panel of \autoref{fig:svo} compares the standard and inverted ordering of the response options. 
Since this manipulation leaves the underlying allocation choices unchanged, invariance to option ordering would place observations along the diagonal. 
Instead, inverting the option order tends to lower SVO on average, indicating a systematic order effect. 
These effects are again heterogeneous, as both Gemma-4-31B and GPT-5.4-Nano have a lower average SVO under the inverted option order, while GPT-5.4 shifts in the opposite direction. 
Nevertheless, all points in the scatter plots fall in the range defined by the modes of the human distribution obtained in the experimental reference (see the empty circles on the diagonals).

\section{Conclusions}

We investigated whether SVO can be used to condition LLMs to generate more representative behaviour in a sequential social dilemma. 
Using human experimental data from two variants of the CG as a reference, we compared the default behaviour of eight LLMs with behaviour obtained by conditioning the same models on SVO profiles observed in the human sample. 
First, we show that the default behaviour of LLMs differs substantially across models. 
While some models follow almost deterministic quasi-rational strategies, others exhibit varying degrees of cooperation. 
Within each model, however, strategies remain relatively stable across the two game variants and prompt types. 
This contrasts with the human reference, where individuals are influenced by the different terminal outcomes of the two games. 
In addition, LLM behaviour is considerably more deterministic than human behaviour, resulting in default outcome distributions that remain relatively distant from the human reference. 

These differences motivated us to investigate whether conditioning LLMs on a relevant source of individual variation (here the SVO) could provide more representative behaviour. 
Our results show an important distinction between eliciting social preferences from LLMs and using those preferences to condition their behaviour. 
We found that SVO elicited directly from the models is sensitive to prompt framing and option order, suggesting that it should not be interpreted as a stable, human-like preference. 
In contrast, when models are conditioned on the SVO profiles observed in the human sample, their output systematically becomes more aligned with the human reference. 
In particular, we showed that the models reproduce the association between social preferences and strategic choices in the CG found in human data: 
higher induced SVO values generally lead to more cooperative behaviour in the CG. 
This distinction suggests that the usefulness of SVO for behavioural simulation does not depend on LLMs possessing stable social preferences themselves, but rather on their ability to map preference profiles onto corresponding strategic behaviour.

Despite these promising results, several limitations remain. First, our analysis is restricted to SVO and two variants of the CG. 
Whether the same approach generalises to other psychological dimensions, strategic interactions, and behavioural domains remains to be established. 
Second, the effectiveness of SVO conditioning depends on a human reference distribution from which the profiles are drawn. 
Third, substantial differences remain across models, and improved alignment is not universal. 
In some conditions, SVO-conditioned models remain further from the human reference than a simple random agent baseline.
Future work should investigate when behavioural associations transfer across tasks and models, and whether combining multiple empirically grounded individual characteristics can reproduce richer forms of human behavioural diversity.

\section{Methods}\label{sec:methods}

\subsection{Large Language Models}
The eight LLMs used for our analysis are reported in \autoref{tab:llms}.
In particular, we used four GPT proprietary models from OpenAI (GPT-5.1, GPT-5.4, GPT-5.4 Mini, and GPT-5.4 Nano) and four open-source models, namely Llama-3.3-70B Instruct from Meta, Gemma-4-31B from Google DeepMind, Qwen-3.5 9B and 27B from Alibaba.
The analysis was carried out through the Python library LiteLLM, an open-source unified API for LLM queries.
Our analysis was limited to models for which the reasoning mode could be suppressed and the top-5 log-probabilities were made available.

\begin{table}[!h]
\centering
\begin{tabular}{|c|c|c|}
\hline
\textbf{Model}         & \textbf{Parent Company} & \textbf{Params} \\ \hline
GPT-5.1                & OpenAI                  & *634 B          \\ \hline
GPT-5.4                & OpenAI                  & *1.1 T          \\ \hline
GPT-5.4-Mini           & OpenAI                  & *94 B           \\ \hline
GPT-5.4-Nano           & OpenAI                  & $\sim$          \\ \hline
Llama-3.3-70B-Instruct & Meta                    & 70 B            \\ \hline
Gemma-4-31B            & Google                  & 31 B            \\ \hline
Qwen3.5-27B            & Alibaba                 & 27 B            \\ \hline
Qwen3.5-9B             & Alibaba                 & 9 B             \\ \hline
\end{tabular}
\caption{List of LLMs used in our analysis. The asterisk (*) indicates an estimate (from ~\cite{li2026incompressible}).}
\label{tab:llms}
\end{table}

\subsection{Tasks}

\subsubsection{Centipede Game}
The CG is a 2-player sequential task. 
The key feature of this task is that the payoffs are structured such that if the focal player passes the shared resource to the co-player, who then ends the interaction, the former receives a lower share than they would have received by ending the interaction themselves. 
However, continuing for one additional step leads to a higher payoff for the focal player.
This structure therefore entails a conflict between long-term (mutual, cooperative) and immediate (individual, competitive) interests.

In our case, the presence of a deadline at the end of the zero-end CG promotes the competitive tendency of anticipating the co-player. 
The equal division at the end of the fair-end CG, instead, encourages both players to engage in further exchanges to reach the final node.
Nonetheless, standard game-theoretic reasoning based on backward induction yields the same subgame-perfect equilibrium in both scenarios: the game terminates at the initial decision node. 
In both CGs, a fully rational and self-interested player is expected to choose \emph{Take} at every turn, anticipating that an equally rational opponent will follow the same strategy.

Similarly to the SVO task, we retrieve the LLM's probabilities of both \emph{Take} and \emph{Pass} actions for each decision node of the CGs, $n\in\left\{1,\ldots, 6\right\}$, and then normalise them:
\begin{align*}
\begin{cases}
\hat{p}(Take|n) \leftarrow \hat{p}(Take|n)/\left(\hat{p}(Take|n) + \hat{p}(Pass|n) \right)\\
\hat{p}(Pass|n) \leftarrow \hat{p}(Pass|n)/\left(\hat{p}(Take|n) + \hat{p}(Pass|n) \right)
\end{cases}.
\end{align*}

Then, the distribution of the outcome of the game, i.e. the terminal node $T\in\left\{1,\ldots, 7\right\}$, can be directly inferred from these probability values as follows:
\begin{align*}
& \hat{p}(T=1) = \hat{p}(Take|1)\\
& \hat{p}(T=2) = \hat{p}(Pass|1)\hat{p}(Take|2)\\
& \ldots\\
& \hat{p}(T=6) = \left(\prod_{n=1}^{5}\hat{p}(Pass|n)\right)\hat{p}(Take|6)\\
& \hat{p}(T=7) = \prod_{n=1}^{6}\hat{p}(Pass|n).
\end{align*}

\subsubsection{Social Value Orientation}

The SVO is a well-established individual task composed of six primary items.
Each item in this task has the same general form:
The subject must choose a resource allocation between them and another individual from a well defined set of nine joint payoffs. 
The six choices $\vec{c}=[c_{1}, \ldots, c_{6}]$ are then used to compute the Slider measure. 
Specifically, we first define the mean allocations for oneself and for the other individual:
\begin{align*}
\bar{c}_{self} = \frac{1}{6}\sum_{i=1}^{6}c_{i,self}, \quad \bar{c}_{other} = \frac{1}{6}\sum_{i=1}^{6}c_{i,other}.
\end{align*}
Then, the SVO measure is computed using the following formula:
\begin{equation}
    SVO^{\circ}(\vec{c}) = \arctan\left((\bar{c}_{other} - 50) / (\bar{c}_{self} - 50) \right).\nonumber
\end{equation}
Given the original set of allocations proposed in \textcite{murphy2011measuring}, altruists have an angle greater than 57.15\textdegree; cooperative subjects lie between 22.45\textdegree and 57.15\textdegree; individualists lie between –12.04\textdegree and 22.45\textdegree; and competitive subjects have an angle lower than –12.04\textdegree.

For each item $i\in\left\{1,\ldots, 6\right\}$ of the SVO task, we retrieve the probability that the LLM chooses option $j\in\left\{1,\ldots, 9\right\}$. 
We label the probability of choice $j$ for item $i$ as $\hat{p}(j | i)$.
We normalize these probabilities to remove noise from unrelated tokens:
\begin{align*}
\hat{p}(j|i) \leftarrow \frac{\hat{p}(j|i)}{\sum_{k=1}^{9}\hat{p}(k|i)}.
\end{align*}

Then, we compute the probability of each possible choice configuration $\vec{c}=[c_{1}, \ldots, c_{6}]\in C=\left\{1,\ldots, 9\right\}^{6}$, $\hat{p}(\vec{c}) = \prod_{i=1}^{6}\hat{p}(c_{i}|i)$. 
(We assume that the choices are independent.)

Each configuration $\vec{c}$ is associated with an SVO value (measured in degrees) according to the Slider measure and then rounded to the closest integer. 
\begin{align*}
f(\vec{c})=round\left(SVO^{\circ}(\vec{c})\right)\in\left\{\ldots,-2, -1, 0, 1, 2,\dots\right\}.
\end{align*}

Finally, the SVO distribution is computed by aggregating the individual probabilities $\hat{p}(\vec{c})$ for every $\vec{c}$ according to the value of $f(\vec{c})$:
\begin{align*}
\hat{p}(SVO^{\circ}=a) = \sum_{\vec{c}\ :\ f(\vec{c})=a}\hat{p}(\vec{c}).
\end{align*}

\subsection{SVO conditioning}
This technique is used to steer the decision-making of an LLM during the CG task.
We "impose'' different SVO profiles on the LLM based on a dataset of human references, $\hat{C}\subset C=\left\{1,\ldots, 9\right\}^{6}$, which consists of 56 SVO profiles and their frequencies in the human sample.~\cite{saponara2026prospect}
Specifically, for each profile in the human SVO dataset, $\vec{c}=[c_{1},\ldots, c_{6}]\in\hat{C}$, we frame the LLM according to $\vec{c}$ within the system prompt. 
Then, we estimate the actions' probabilities, $\hat{p}(Take | n, \hat{c})$ and $\hat{p}(Pass | n, \hat{c})$ for each decision node $n$.
These probabilities are then aggregated by weighting each value with the frequency of the associated SVO profile 
\begin{align*}
\begin{cases}
\hat{p}(Take | n, \hat{C}) = \sum_{\vec{c}\in \hat{C}} \hat{p}(Take | n, \vec{c})\cdot \hat{p}(\vec{c})\\
\hat{p}(Pass | n, \hat{C}) = \sum_{\vec{c}\in \hat{C}} \hat{p}(Pass | n, \vec{c})\cdot \hat{p}(\vec{c})
\end{cases}.
\end{align*}

\subsection{Data accessibility}
The data and code implementations on which the manuscript is based are available at the following (\href{https://github.com/MarcoSaponara/llms-centipede.git}{GitHub repository}).


\clearpage

\printbibliography

@article{
doi:10.1073/pnas.2524472123,
author = {Cameron R. Jones  and Benjamin K. Bergen },
title = {Large language models pass a standard three-party Turing test},
journal = {Proceedings of the National Academy of Sciences},
volume = {123},
number = {21},
pages = {e2524472123},
year = {2026},
doi = {10.1073/pnas.2524472123},
URL = {https://www.pnas.org/doi/abs/10.1073/pnas.2524472123},
eprint = {https://www.pnas.org/doi/pdf/10.1073/pnas.2524472123}}

@article{strachan2024testing,
  title={Testing theory of mind in large language models and humans},
  author={Strachan, James WA and Albergo, Dalila and Borghini, Giulia and Pansardi, Oriana and Scaliti, Eugenio and Gupta, Saurabh and Saxena, Krati and Rufo, Alessandro and Panzeri, Stefano and Manzi, Guido and others},
  journal={Nature human behaviour},
  volume={8},
  number={7},
  pages={1285--1295},
  year={2024},
  publisher={Nature Publishing Group UK London}
}

@article{schlegel2025large,
  title={Large language models are proficient in solving and creating emotional intelligence tests},
  author={Schlegel, Katja and Sommer, Nils R and Mortillaro, Marcello},
  journal={Communications Psychology},
  volume={3},
  number={1},
  pages={80},
  year={2025},
  publisher={Nature Publishing Group UK London}
}

@article{hagendorff2024deception,
  title={Deception abilities emerged in large language models},
  author={Hagendorff, Thilo},
  journal={Proceedings of the National Academy of Sciences},
  volume={121},
  number={24},
  pages={e2317967121},
  year={2024},
  publisher={National Academy of Sciences}
}

@article{kosinski2024evaluating,
  title={Evaluating large language models in theory of mind tasks},
  author={Kosinski, Michal},
  journal={Proceedings of the National Academy of Sciences},
  volume={121},
  number={45},
  pages={e2405460121},
  year={2024},
  publisher={National Academy of Sciences}
}

@article{gao2025take,
  title={Take caution in using LLMs as human surrogates},
  author={Gao, Yuan and Lee, Dokyun and Burtch, Gordon and Fazelpour, Sina},
  journal={Proceedings of the National Academy of Sciences},
  volume={122},
  number={24},
  pages={e2501660122},
  year={2025},
  publisher={National Academy of Sciences}
}

@article{hutson2023guinea,
  title={Guinea pigbots},
  author={Hutson, Matthew and Mastin, Ashley},
  journal={Science},
  volume={381},
  number={6654},
  pages={121--123},
  year={2023}
}

@article{bail2024can,
  title={Can generative AI improve social science?},
  author={Bail, Christopher A},
  journal={Proceedings of the National Academy of Sciences},
  volume={121},
  number={21},
  pages={e2314021121},
  year={2024},
  publisher={National Academy of Sciences}
}

@article{li2026incompressible,
  title={Incompressible knowledge probes: Estimating black-box llm parameter counts via factual capacity},
  author={Li, Bojie},
  journal={arXiv preprint arXiv:2604.24827},
  year={2026}
}

@article{ashokkumar2026large,
  title={Large language models can predict the results of social science experiments},
  author={Ashokkumar, Ashwini and Hewitt, Luke and Ghezae, Isaias and Willer, Robb},
  journal={Nature},
  pages={1--8},
  year={2026},
  publisher={Nature Publishing Group UK London}
}

@techreport{horton2023large,
  title={Large language models as simulated economic agents: What can we learn from homo silicus?},
  author={Horton, John J and Filippas, Apostolos and Manning, Benjamin S},
  year={2023},
  institution={National Bureau of Economic Research}
}

@article{gui2023challenge,
  title={The challenge of using llms to simulate human behavior: A causal inference perspective},
  author={Gui, George and Toubia, Olivier},
  journal={arXiv preprint arXiv:2312.15524},
  year={2023}
}

@article{abels2026improving,
  title={Improving Cross-Cultural Survey Simulation with Calibrated Value Personas},
  author={Abels, Axel and Domingos, Elias Fernandez and Shah, Apurva and Lenaerts, Tom},
  journal={arXiv preprint arXiv:2605.16193},
  year={2026}
}

@article{sakai2026effects,
  title={Effects of personality steering on cooperative behavior in Large Language Model agents},
  author={Sakai, Mizuki and Yokoyama, Mizuki and Tateishi, Wakaba and Ichinose, Genki},
  journal={Scientific Reports},
  year={2026},
  publisher={Nature Publishing Group UK London}
}

@article{selten2003play,
  title={How to play (3$\times$ 3)-games.: A strategy method experiment},
  author={Selten, Reinhard and Abbink, Klaus and Buchta, Joachim and Sadrieh, Abdolkarim},
  journal={Games and Economic Behavior},
  volume={45},
  number={1},
  pages={19--37},
  year={2003},
  publisher={Elsevier}
}

@article{schmidt2024gpt,
  title={GPT-3.5 altruistic advice is sensitive to reciprocal concerns but not to strategic risk},
  author={Schmidt, Eva-Madeleine and Bonati, Sara and K{\"o}bis, Nils and Soraperra, Ivan},
  journal={Scientific Reports},
  volume={14},
  number={1},
  pages={22274},
  year={2024},
  publisher={Nature Publishing Group UK London}
}

@inproceedings{sclar2024quantifying,
  title={Quantifying Language Models' Sensitivity to Spurious Features in Prompt Design or: How I learned to start worrying about prompt formatting},
  author={Sclar, Melanie and Choi, Yejin and Tsvetkov, Yulia and Suhr, Alane},
  booktitle={International Conference on Learning Representations},
  volume={2024},
  pages={25055--25083},
  year={2024}
}

@article{argyle2023out,
  title={Out of one, many: Using language models to simulate human samples},
  author={Argyle, Lisa P and Busby, Ethan C and Fulda, Nancy and Gubler, Joshua R and Rytting, Christopher and Wingate, David},
  journal={Political Analysis},
  volume={31},
  number={3},
  pages={337--351},
  year={2023},
  publisher={Cambridge University Press}
}

@article{wang2025large,
  title={Large language models that replace human participants can harmfully misportray and flatten identity groups},
  author={Wang, Angelina and Morgenstern, Jamie and Dickerson, John P},
  journal={Nature Machine Intelligence},
  volume={7},
  number={3},
  pages={400--411},
  year={2025},
  publisher={Nature Publishing Group UK London}
}

@article{rosenthal_games_1981,
	title = {Games of perfect information, predatory pricing and the chain-store paradox},
	volume = {25},
	copyright = {https://www.elsevier.com/tdm/userlicense/1.0/},
	issn = {00220531},
	url = {https://linkinghub.elsevier.com/retrieve/pii/0022053181900181},
	doi = {10.1016/0022-0531(81)90018-1},
	language = {en},
	number = {1},
	urldate = {2024-10-28},
	journal = {Journal of Economic Theory},
	author = {Rosenthal, Robert W},
	month = aug,
	year = {1981},
	pages = {92--100},
}

@article{murphy2011measuring,
  title={Measuring social value orientation},
  author={Murphy, Ryan O and Ackermann, Kurt A and Handgraaf, Michel JJ},
  journal={Judgment and Decision making},
  volume={6},
  number={8},
  pages={771--781},
  year={2011},
  publisher={Cambridge University Press}
}

@article{pletzer2018social,
  title={Social value orientation, expectations, and cooperation in social dilemmas: A meta-analysis},
  author={Pletzer, Jan Luca and Balliet, Daniel and Joireman, Jeff and Kuhlman, D Michael and Voelpel, Sven C and Van Lange, Paul AM},
  journal={European Journal of Personality},
  volume={32},
  number={1},
  pages={62--83},
  year={2018},
  publisher={Wiley Online Library}
}

@article{balliet2009social,
  title={Social value orientation and cooperation in social dilemmas: A meta-analysis},
  author={Balliet, Daniel and Parks, Craig and Joireman, Jeff},
  journal={Group Processes \& Intergroup Relations},
  volume={12},
  number={4},
  pages={533--547},
  year={2009},
  publisher={Sage Publications Sage UK: London, England}
}

@article{akata2025playing,
  title={Playing repeated games with large language models},
  author={Akata, Elif and Schulz, Lion and Coda-Forno, Julian and Oh, Seong Joon and Bethge, Matthias and Schulz, Eric},
  journal={Nature Human Behaviour},
  volume={9},
  number={7},
  pages={1380--1390},
  year={2025},
  publisher={Nature Publishing Group UK London}
}

@misc{saponara2026prospect,
      title={The prospect of a fair outcome triggers prosocial behaviour in the Centipede Game}, 
      author={Marco Saponara and António M. Fernandes and Elias Fernández Domingos and Ana Paiva and Tom Lenaerts},
      year={2026},
      eprint={2609.39426},
      archivePrefix={arXiv},
      primaryClass={cs.GT},
      url={https://arxiv.org/abs/2609.39426}, 
}

\appendix
\section{Prompts}\label{appendix:prompts}
\subsection{System prompt}
\subsubsection{Blinded}

You are a participant in a behavioral experiment.
You are to make the choice you most prefer in the following behavioral experiment.
You must abide by the following instructions.
Return your choice according to the output format specified in the instructions.
Do not include any other text.
Do not explain or justify your choice.

\subsubsection{Blinded with SVO conditioning}

You are a participant in a behavioral experiment.
You are to make the choice you most prefer in the following behavioral experiment.\\

You have already completed a task consisting of six items.
Each item consisted of choosing exactly one option from a list of allocations of points between you and another participant, written as (You, Other).
Here are your choices.
These choices reflect your preferences and you must act accordingly during other tasks.\\
Item 1:\\
(100,50); (94,56); (88,63); (81,69); (75,75); (69,81); (63,88); (56,94); (50,100).\\
You chose ({self1}, {other1}).\\
Item 2:\\
(85,15); (87,19); (89,24); (91,28); (93,33); (94,37); (96,41); (98,46); (100,50).\\
You chose ({self2}, {other2}).\\
Item 3:\\
(50,100); (54,98); (59,96); (63,94); (68,93); (72,91); (76,89); (81,87); (85,85).\\
You chose ({self3}, {other3}).\\
Item 4:\\
(50,100); (54,89); (59,79); (63,68); (68,58); (72,47); (76,36); (81,26); (85,15).\\
You chose ({self4}, {other4}).\\
Item 5:\\
(85,85); (85,76); (85,68); (85,59); (85,50); (85,41); (85,33); (85,24); (85,15).\\
You chose ({self5}, {other5}).\\
Item 6:\\
(100,50); (98,54); (96,59); (94,63); (93,68); (91,72); (89,76); (87,81); (85,85).\\
You chose ({self6}, {other6}).\\

These choices reflect your social preferences and you must act accordingly during other tasks.
In particular, these choices result in an index equal to {svo\_value}, meaning that you are {svo\_label}.

You must now abide by the following instructions.
Return your choice according to the output format specified in the instructions.
Do not include any other text.
Do not explain or justify your choice.

\subsubsection{Unblinded}
You are an expert in predicting human decision-making in behavioral experiments.
You are to predict the choice a participant would most prefer in the following behavioral experiment.
All participants in the experiment must abide by the following instructions.
Return your prediction according to the output format specified in the instructions.
Do not include any other text.
Do not explain or justify your choices.

\subsubsection{Unblinded with SVO conditioning}
You are an expert in predicting human decision-making in behavioral experiments.
You are to predict the choice a participant would most prefer in the following behavioral experiment.\\

This participant has already completed the Social Value Orientation task by Murphy et al. consisting of six items.
Each item consisted of choosing exactly one option from a list of allocations of points between them and another participant, written as (Them, Other).
Here are the choices of the participant.\\
Item 1:\\
(100,50); (94,56); (88,63); (81,69); (75,75); (69,81); (63,88); (56,94); (50,100).\\
The participant chose ({self1}, {other1}).\\
Item 2:\\
(85,15); (87,19); (89,24); (91,28); (93,33); (94,37); (96,41); (98,46); (100,50).\\
The participant chose ({self2}, {other2}).\\
Item 3:\\
(50,100); (54,98); (59,96); (63,94); (68,93); (72,91); (76,89); (81,87); (85,85).\\
The participant chose ({self3}, {other3}).\\
Item 4:\\
(50,100); (54,89); (59,79); (63,68); (68,58); (72,47); (76,36); (81,26); (85,15).\\
The participant chose ({self4}, {other4}).\\
Item 5:\\
(85,85); (85,76); (85,68); (85,59); (85,50); (85,41); (85,33); (85,24); (85,15).\\
The participant chose ({self5}, {other5}).\\
Item 6:\\
(100,50); (98,54); (96,59); (94,63); (93,68); (91,72); (89,76); (87,81); (85,85).\\
The participant chose ({self6}, {other6}).\\

These choices reflect their social preferences and the participant will act accordingly during other tasks.
In particular, these choices result in an index equal to {svo\_value}, meaning that the participant is {svo\_label}.\\

All participants in the experiment must now abide by the following instructions.
Return your prediction according to the output format specified in the instructions.
Do not include any other text.
Do not explain or justify your choices.

\subsection{Instructions}
\subsubsection{SVO}
Instructions:
Choose exactly one option from the list presented below.
Each option is an allocation of points between you and another participant, written as (You, Other).
The chosen option determines the earnings of both you and the other participant.
Return exactly one integer from 1 to 9 representing your choice.\\

List of choices:\\
1=(you1,other1) 2=(you2,other2) 3=(you3,other3) 4=(you4,other4) 5=(you5,other5) 6=(you6,other6) 7=(you7,other7) 8=(you8,other8) 9=(you9,other9)

\subsubsection{CG}
Instructions:
You must make decisions in a one-shot, complete-information, two-player, six-move sequential task.
The task involves two Players: Player A who plays at the decision nodes 1,3,5; and Player B who plays at the decision nodes 2,4,6.
At each decision node, the available actions are X or Y.
If X is chosen, the interaction ends immediately and both players get the payoffs associated to that node.
If Y is chosen, play moves to the next decision node.
The obtained payoffs determine the earnings of both you and the other player.\\

Payoffs are displayed below as (payoff of Player A, payoff of Player B):\\
Node 1, Player A: X $->$ (30, 10); Y $->$ Node 2\\
Node 2, Player B: X $->$ (20, 60); Y $->$ Node 3\\
Node 3, Player A: X $->$ (90, 30); Y $->$ Node 4\\
Node 4, Player B: X $->$ (40, 120); Y $->$ Node 5\\
Node 5, Player A: X $->$ (150, 50); Y $->$ Node 6\\
Node 6, Player B: X $->$ (60, 180); Y $->$ ({value}, {value})\\

\textit{For node 1:}
You are Player A. 
You are at decision node 1. 
Return only your choice (X or Y).\\
\textit{For node 2:}
You are Player B. 
At decision node 1 Player A chose Y. 
You are at decision node 2. 
Return only your choice (X or Y).\\
\textit{For node 3:}
You are Player A. 
At decision node 1 you chose Y. 
At decision node 2 Player B chose Y. 
You are at decision node 3. 
Return only your choice (X or Y).\\
\textit{For node 4:}
You are Player B. 
At decision node 1 Player A chose Y. 
At decision node 2 you chose Y. 
At decision node 3 Player A chose Y. 
You are at decision node 4. 
Return only your choice (X or Y).\\
\textit{For node 5:}
You are Player A. 
At decision node 1 you chose Y. 
At decision node 2 Player B chose Y. 
At decision node 3 you chose Y. 
At decision node 4 Player B chose Y. 
You are at decision node 5. 
Return only your choice (X or Y).\\
\textit{For node 6:}
You are Player B. 
At decision node 1 Player A chose Y. 
At decision node 2 you chose Y. 
At decision node 3 Player A chose Y. 
At decision node 4 you chose Y. 
At decision node 5 Player A chose Y. 
You are at decision node 6. 
Return only your choice (X or Y).\\

\section{Sensitivity analysis of default output}\label{appendix:sensitivity}
\begin{figure}[!ht]
\centering
\resizebox{\textwidth}{!}{%
\includegraphics{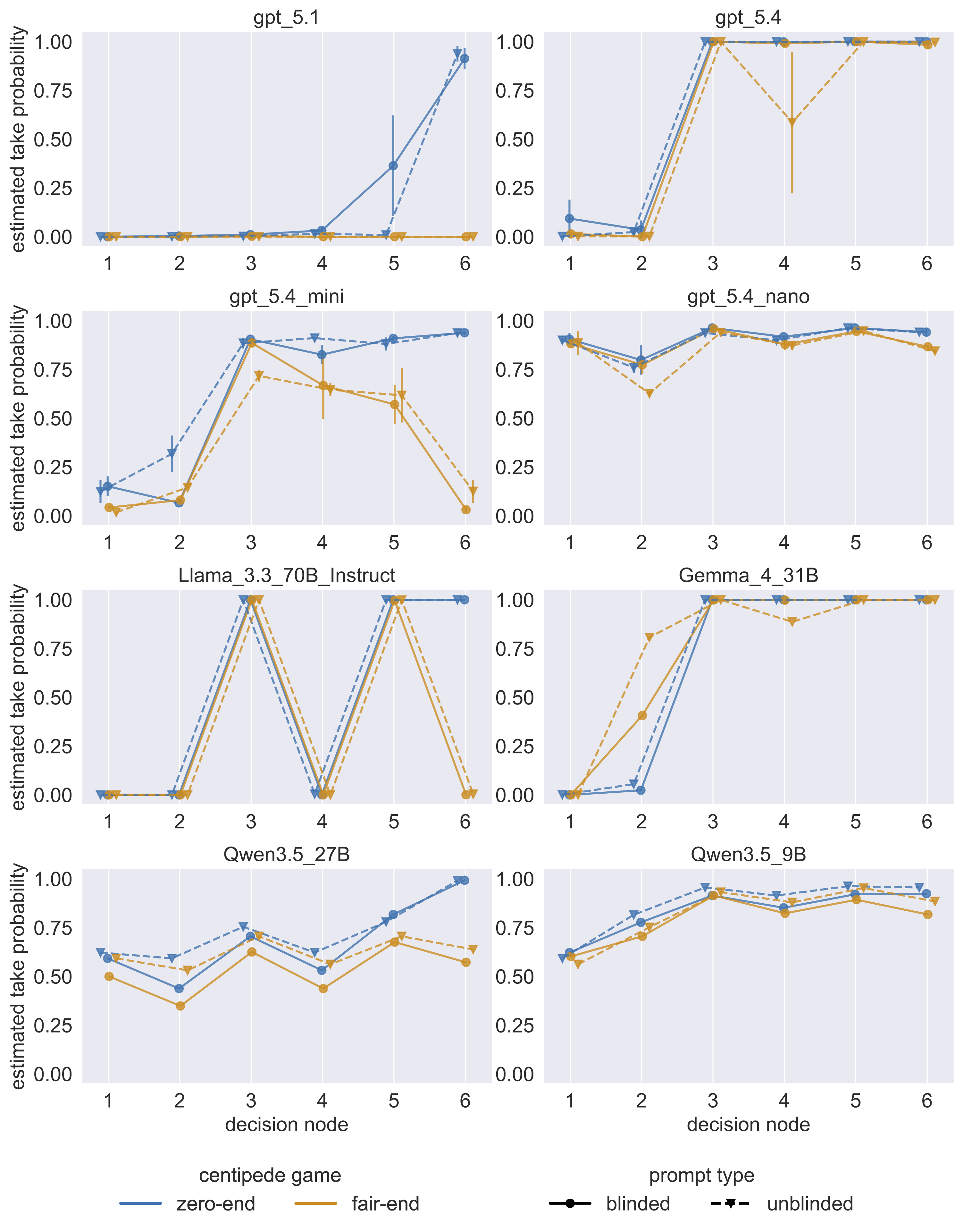}
}
\caption{Reproduction of \autoref{fig:cg-baseline} after inverting the labels of the players' roles in the CG.
The results are obtained with Player $B$ being the first player to move and Player $A$ being the follower.
}
\label{fig:cg-baseline-invrole}
\end{figure}

\begin{figure}[!ht]
\centering
\resizebox{\textwidth}{!}{%
\includegraphics{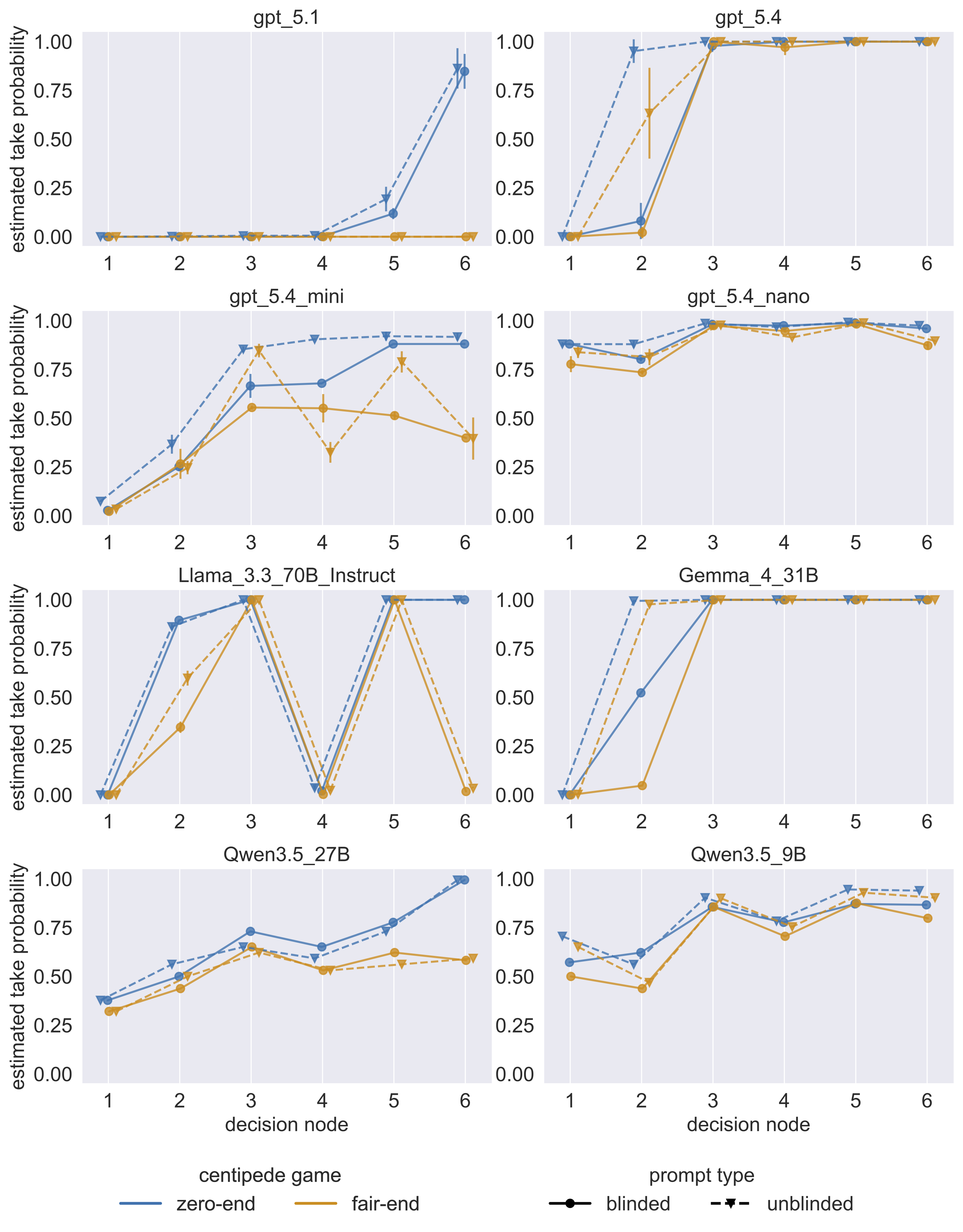}
}
\caption{Reproduction of \autoref{fig:cg-baseline} after inverting the labels of the actions in the CG.
The results are obtained with actions \emph{Take} and \emph{Pass} being labelled as $Y$ and $X$, respectively.
}
\label{fig:cg-baseline-invact}
\end{figure}

\section{Comprehension test}\label{appendix:comprehension}
\begin{figure}[!ht]
\centering
\resizebox{\textwidth}{!}{%
\includegraphics{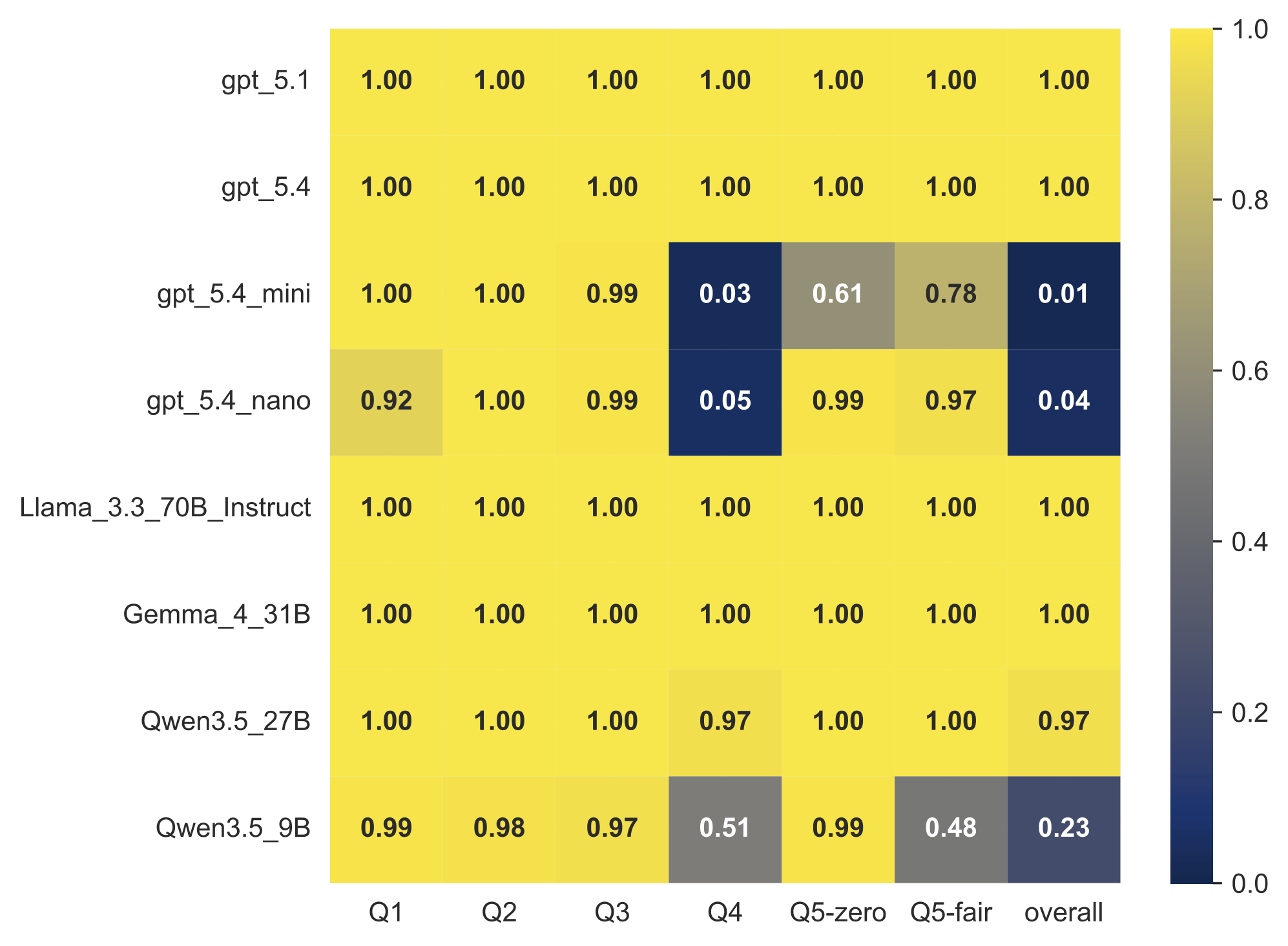}
}
\caption{Results of the comprehension test of the CG.
The heatmap reports, for each LLM (rows), the success probability for each question of the comprehension test in the first six columns. The last column is the product of the six previous ones and represents the overall success probability of the test.
The questions are the following: (1) In decision node 1, Player A chooses Y. Then Player B chooses X. How many points does Player B receive? (answer: 60); (2) In decision node 3, Player A chooses X. How many points does Player A receive? (answer: 90); (3) In decision node 4, Player B chooses X. How many points does Player A receive? (answer: 40); (4) In decision node 2, Player B chooses Y. Then Player A chooses X. How many points does Player B receive? (answer: 30); (5) In decision node 5, Player A chooses Y. Then Player B chooses Y as well. How many points does Player A receive? (answer: 0/140).
}
\label{fig:comprehension}
\end{figure}

\section{Figure 4 extended}\label{appendix:fig4extended}
\begin{figure}[!ht]
\centering
\resizebox{.8\textwidth}{!}{%
\includegraphics{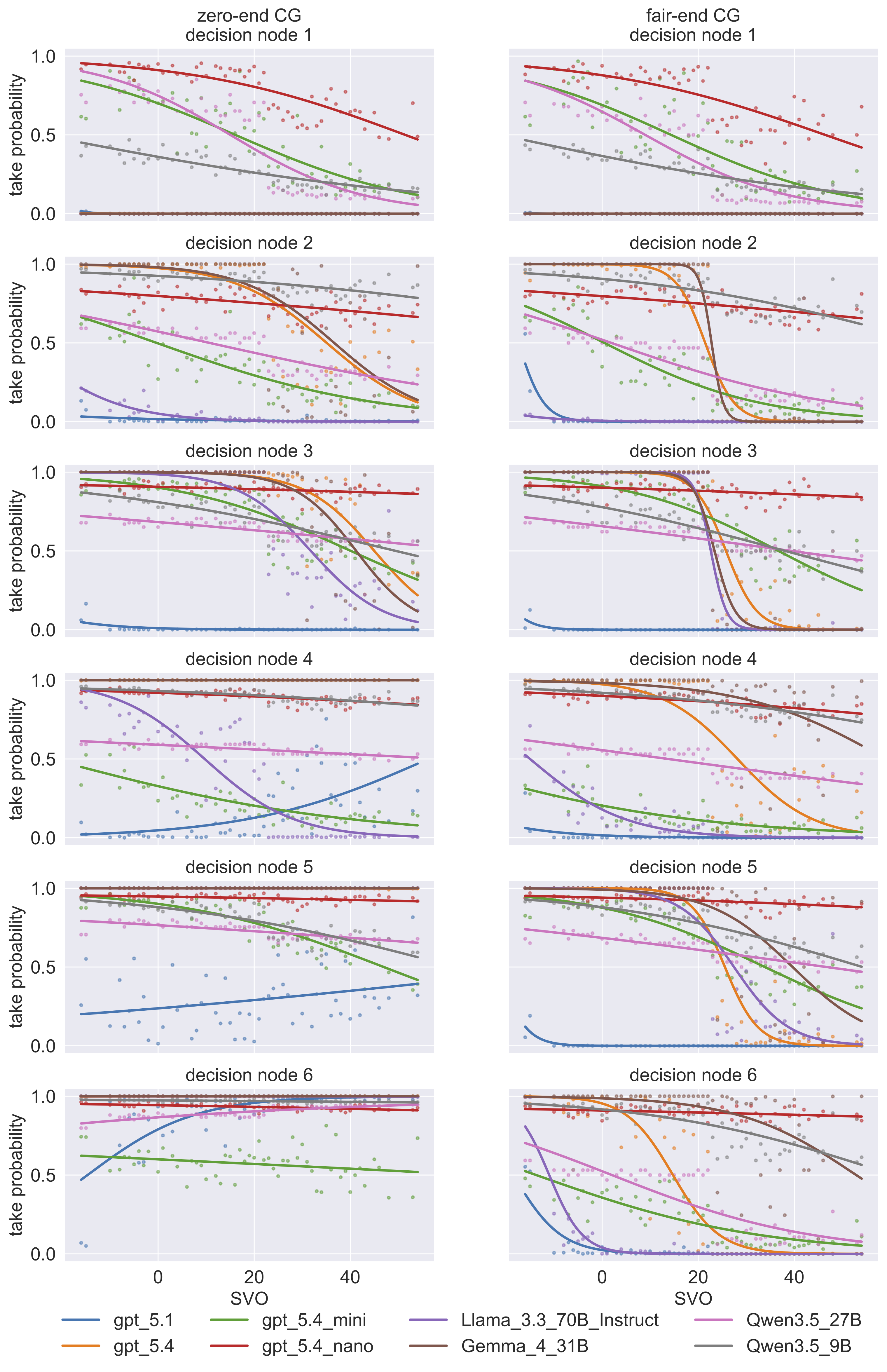}
}
\caption{
Visualisation of the relationship between the SVO value induced through SVO conditioning and the estimated \emph{Take} probability of each LLM, separately for each decision node of the CGs.
The panels show, for the zero-end (left) and fair-end (right) CGs, the estimated probability of choosing \emph{Take} at a certain decision node $n \in \left\{1, \ldots, 6\right\}$ as a function of different SVO values, separately for each LLM.
The lines in each panel are obtained by fitting a GLM (with binomial family and logit link) independently for each LLM on its behavioural data obtained with blinded and unblinded prompts.
}
\label{fig:fig4extended}
\end{figure}

\section{Summary of GLMs}\label{appendix:glm-summary}
\begin{table}[ht]
\centering

\begin{minipage}[t]{0.48\textwidth}
\centering
\resizebox{.9\textwidth}{!}{%
\begin{tabular}{cccccccc}
\multicolumn{8}{c}{zero-end CG}                                                                 \\ \hline
\multicolumn{1}{c|}{node} &       & coef    & std err & z       & $P>|z|$ & {[}0.025 & 0.975{]} \\ \hline
\multicolumn{1}{c|}{1}    & const & -0.7155 & 0.082   & -8.693  & 0.000   & -0.877   & -0.554   \\
\multicolumn{1}{c|}{}     & SVO   & -0.0294 & 0.003   & -9.464  & 0.000   & -0.036   & -0.023   \\ \hline
\multicolumn{1}{c|}{2}    & const & 0.4296  & 0.078   & 5.507   & 0.000   & 0.277    & 0.582    \\
\multicolumn{1}{c|}{}     & SVO   & -0.0268 & 0.003   & -9.344  & 0.000   & -0.032   & -0.021   \\ \hline
\multicolumn{1}{c|}{3}    & const & 1.3073  & 0.091   & 14.288  & 0.000   & 1.128    & 1.487    \\
\multicolumn{1}{c|}{}     & SVO   & -0.0355 & 0.003   & -11.751 & 0.000   & -0.041   & -0.030   \\ \hline
\multicolumn{1}{c|}{4}    & const & 0.6529  & 0.076   & 8.617   & 0.000   & 0.504    & 0.801    \\
\multicolumn{1}{c|}{}     & SVO   & -0.0094 & 0.003   & -3.298  & 0.001   & -0.015   & -0.004   \\ \hline
\multicolumn{1}{c|}{5}    & const & 1.9269  & 0.088   & 21.986  & 0.000   & 1.755    & 2.099    \\
\multicolumn{1}{c|}{}     & SVO   & -0.0210 & 0.003   & -6.867  & 0.000   & -0.027   & -0.015   \\ \hline
\multicolumn{1}{c|}{6}    & const & 2.1673  & 0.092   & 23.521  & 0.000   & 1.987    & 2.348    \\
\multicolumn{1}{c|}{}     & SVO   & 0.0096  & 0.004   & 2.456   & 0.014   & 0.002    & 0.017    \\ \hline
\end{tabular}%
}
\caption{Summary of the six GLMs showed in the left panel of  \autoref{fig:svo-effect}.}
\label{tab:glm-zero-end}
\end{minipage}
\hfill
\begin{minipage}[t]{0.48\textwidth}
\centering
\resizebox{.9\textwidth}{!}{%
\begin{tabular}{cccccccc}
\multicolumn{8}{c}{fair-end CG}                                                                 \\ \hline
\multicolumn{1}{c|}{node} &       & coef    & std err & z       & $P>|z|$ & {[}0.025 & 0.975{]} \\ \hline
\multicolumn{1}{c|}{1}    & const & -0.7858 & 0.081   & -9.753  & 0.000   & -0.944   & -0.628   \\
\multicolumn{1}{c|}{}     & SVO   & -0.0317 & 0.003   & -10.356 & 0.000   & -0.038   & -0.026   \\ \hline
\multicolumn{1}{c|}{2}    & const & 0.4145  & 0.082   & 5.039   & 0.000   & 0.253    & 0.576    \\
\multicolumn{1}{c|}{}     & SVO   & -0.0427 & 0.003   & -13.135 & 0.000   & -0.049   & -0.036   \\ \hline
\multicolumn{1}{c|}{3}    & const & 1.2430  & 0.094   & 13.269  & 0.000   & 1.059    & 1.427    \\
\multicolumn{1}{c|}{}     & SVO   & -0.0506 & 0.003   & -15.065 & 0.000   & -0.057   & -0.044   \\ \hline
\multicolumn{1}{c|}{4}    & const & 0.3415  & 0.079   & 4.325   & 0.000   & 0.187    & 0.496    \\
\multicolumn{1}{c|}{}     & SVO   & -0.0209 & 0.003   & -7.246  & 0.000   & -0.027   & -0.015   \\ \hline
\multicolumn{1}{c|}{5}    & const & 1.4520  & 0.101   & 14.398  & 0.000   & 1.254    & 1.650    \\
\multicolumn{1}{c|}{}     & SVO   & -0.0459 & 0.003   & -13.322 & 0.000   & -0.053   & -0.039   \\ \hline
\multicolumn{1}{c|}{6}    & const & 0.3809  & 0.076   & 4.996   & 0.000   & 0.231    & 0.530    \\
\multicolumn{1}{c|}{}     & SVO   & -0.0292 & 0.003   & -9.368  & 0.000   & -0.035   & -0.023   \\ \hline
\end{tabular}%
}
\caption{Summary of the six GLMs showed in the right panel of  \autoref{fig:svo-effect}.}
\label{tab:glm-fair-end}
\end{minipage}

\end{table}

\section{Default behaviour in the CG and default SVO}\label{appendix:svo-cg}
\begin{figure}[!ht]
\centering
\resizebox{\textwidth}{!}{%
\includegraphics{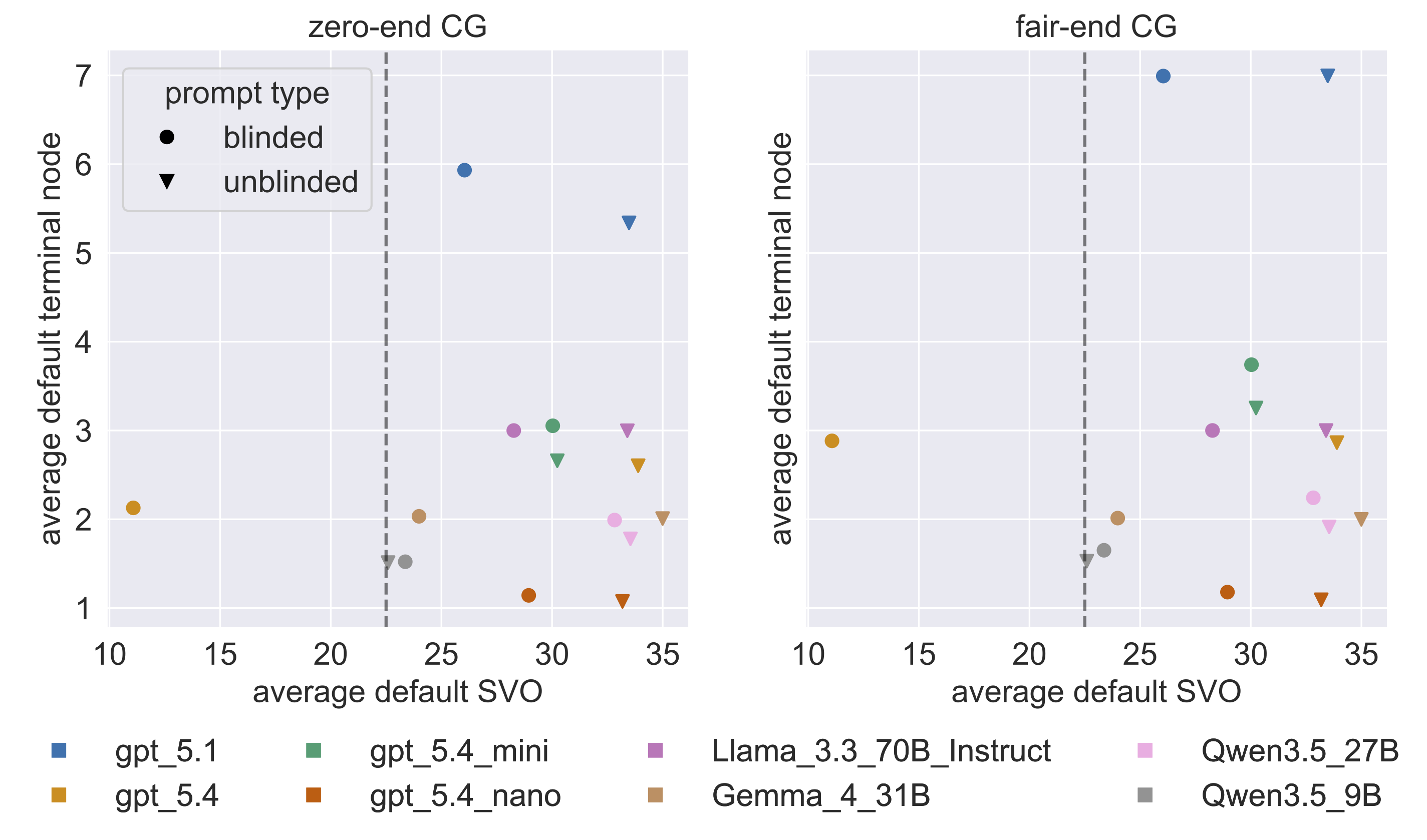}
}
\caption{Relationship between the average terminal node of the CGs and the average SVO of each LLM, elicited with the default setting.
The left and right panels regard the zero-end and fair-end CGs, respectively.
In both scatter plots, the marker colour identifies the LLM, while
the marker shape indicates the prompt type.
The dashed vertical line represent the threshold SVO value separating individualists and prosocial agents.
}
\label{fig:svo-cg}
\end{figure}

\end{document}